\documentclass[twocolumn,pre,showpacs,superscriptaddress]{revtex4}
\usepackage{graphicx}%
\usepackage{color} 
\usepackage{transparent}
\usepackage{psfrag}
\usepackage{dcolumn}
\usepackage{amsmath}
\usepackage{amssymb}
\usepackage{latexsym}
\usepackage{hyperref} 
\usepackage{booktabs}
\usepackage{latexsym}
\usepackage{natbib}
\begin{document}

\def\bea{\begin{eqnarray}}
\def\eea{\end{eqnarray}}
\def\beq{\begin{equation}}
\def\eeq{\end{equation}}
\def\f{\frac}
\def\k{\kappa}
\def\e{\epsilon}
\def\ve{\varepsilon}
\def\be{\beta}
\def\D{\Delta}
\def\h{\theta}
\def\t{\tau}
\def\a{\alpha}

\def\cDa{{\cal D}[X]}
\def\cD{{\cal D}[x]}
\def\cL{{\cal L}}
\def\cLo{{\cal L}_0}
\def\cLa{{\cal L}_1}

\def\Re{{\rm Re}}
\def\sj{\sum_{j=1}^2}
\def\rk{\rho^{ (k) }}
\def\rek{\rho^{ (1) }}
\def\cek{C^{ (1) }}
\def\rz{\rho^{ (0) }}
\def\rt{\rho^{ (2) }}
\def\rtb{\bar \rho^{ (2) }}
\def\trk{\tilde\rho^{ (k) }}
\def\trek{\tilde\rho^{ (1) }}
\def\trz{\tilde\rho^{ (0) }}
\def\trt{\tilde\rho^{ (2) }}
\def\r{\rho}
\def\tD{\tilde {D}}

\def\s{\sigma}
\def\kb{k_B}
\def\F{{\cal F}}
\def\la{\langle}
\def\ra{\rangle}
\def\nn{\nonumber}
\def\up{\uparrow}
\def\dn{\downarrow}
\def\S{\Sigma}
\def\dg{\dagger}
\def\d{\delta}
\def\p{\partial}
\def\l{\lambda}
\def\L{\Lambda}
\def\G{\Gamma}
\def\o{\Omega}
\def\w{\omega}
\def\g{\gamma}

\def\jv{ {\bf j}}
\def\jr{ {\bf j}_r}
\def\jd{ {\bf j}_d}
\def\noi{\noindent}
\def\a{\alpha}
\def\d{\delta}
\def\p{\partial} 

\def\la{\langle}
\def\ra{\rangle}
\def\e{\epsilon}
\def\n{\eta}
\def\g{\gamma}
\def\break#1{\pagebreak \vspace*{#1}}
\def\hf{\frac{1}{2}}

\title{ Absence of jamming in ant trails: Feedback control of self propulsion and noise} 
\author{Debasish Chaudhuri}
\email{debc@iith.ac.in}
\affiliation{Indian Institute of Technology Hyderabad,
Yeddumailaram 502205, Telengana, India
}
\author{Apoorva Nagar}
\email{apoorva.nagar@iist.ac.in}
\affiliation{Indian Institute of Space Science and Technology,
Thiruvananthapuram, Kerala, India
}

\date{\today}

\begin{abstract}
We present a model of ant traffic considering individual ants as self-propelled particles undergoing single file motion on a one-dimensional trail. 
Recent experiments on unidirectional ant traffic in well-formed natural trails showed that the collective velocity of ants remains approximately unchanged, leading to absence of jamming even at very high densities~[John {\em et. al.}, Phys. Rev. Lett. {\bf 102}, 108001 (2009)]. Assuming a feedback control mechanism of self-propulsion force generated by each ant using information about the distance from the ant in front, our model captures all the main features observed in the experiment. The distance headway distribution shows a maximum corresponding to separations within clusters. The position of this maximum remains independent of average number density. We find a non-equilibrium first order transition, with the formation of an {\em infinite cluster} at a threshold density where all the ants in the system suddenly become part of a single cluster.

\end{abstract}

\maketitle

\section{Introduction}
The study of collective motion of self propelled particles 
-- from sub-cellular machines like molecular motors moving on polymeric tracks to unicellular life forms like bacteria, from the co-ordinated motion of insects as small as ants to large mammals like humans -- shows  emergence of rich dynamical behavior and patterns starting with simple rules for the motion of individual 
units~\cite{Vicsek2012,Chou2011,romanczuk2012active}. The study of ants, in particular, is fascinating from more than one perspective~\cite{Hoelldobler1990}. 
From a traffic point of view, the collective motion of ants shows 
self organization of flow to maximize efficiency in transport~\cite{Dussutour2005,Dussutour2004}, and spontaneous formation of lanes in bi-directional traffic~\cite{Fourcassie2010,Couzin2003}. Another interesting feature of ant motion is the spontaneous selection of shortest path between the nest and the food source by using only local dynamical rules, without the aid of a global perspective. This has inspired theoretical work on new kinds of optimization algorithms~\cite{Blum2005,Dorigo2005}. While walking, ants leave chemical trails in form of pheromone, that later ants follow leading to ant trail formation~\cite{Hoelldobler1990,Ganeshaiah1991, Millonas1992, Watmough1995,S.Camazine2001,Chialvo1995, Rauch1995}. Formation of these trails have been described theoretically in terms of active-walker models. The mechanism is ubiquitous in nature and similar to river basin formation~\cite{Scheidegger1967}, formation of pedestrian trails~\cite{Helbing1997, Helbing1997a}, and formation of axon bundles in mammalian sensory neurons~\cite{Chaudhuri2009c, Chaudhuri2011b}. 

\begin{figure}[t]
\begin{center}
\includegraphics[width=8cm]{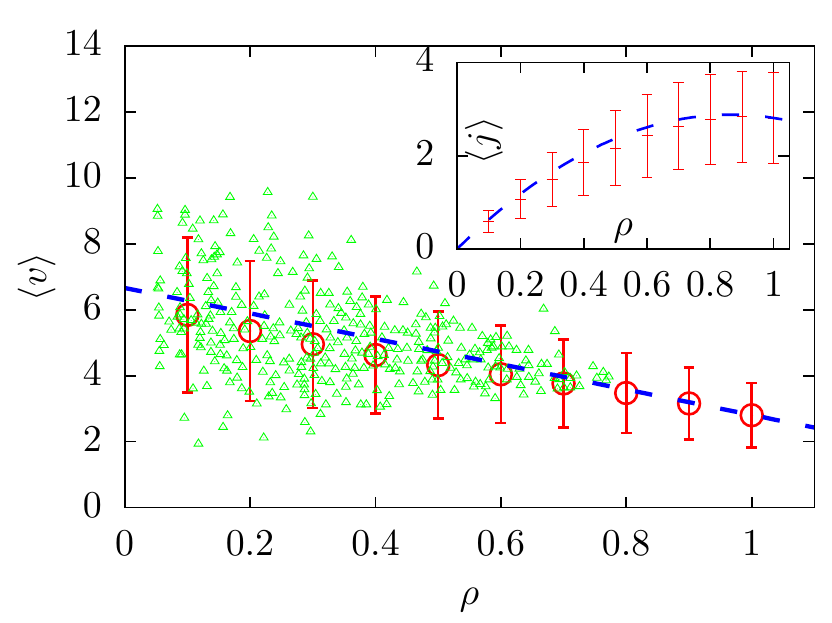}
\caption{(Color online) Average velocity with density. The symbols $\circ$ with error bars denote our simulation results, while the
small $\triangle$ symbols denote data extracted from Fig.3 of Ref.~\cite{John2009}.  The dashed line is the mean filed estimate $\la v \ra = f_0 (1-\r/\r_f)$. 
Inset: The { fundamental diagram} showing current as a function of density. 
The dashed line is a plot of mean field estimate $\la j \ra = \r \la v \ra$.}
\label{mean_v}
\end{center}
\end{figure}

A recent experimental study on collective motion of ants within preformed natural trails in the species {\em Leptogenys processionalis} showed several intriguing features~\cite{John2009}. It found absence of  jamming of ant-traffic even at very high densities -- with only a minor decrease in velocity at higher density, reduction in velocity-fluctuations with increasing densities, as well as formation of clusters of ants within the trail. The flow behavior is in contrast to vehicular traffic where a decrease of flux is observed at high densities, indicating congestion and a tendency to form jams, captured by flux-density plots known as fundamental diagrams of traffic flow~\cite{Kerner2009}. 
In this paper, we present a model of self-propelled particles performing  single file motion, which captures all the main observations of Ref.~\cite{John2009}.
In single file motion particles constrained to move in one dimension can not overtake each other, performing sub-diffusive dynamics~\cite{Hodgkin1955, Lizana2008, Barkai2009, Hahn1996,Wei2000}. 

 Earlier theoretical work on ant traffic using asymmetric hopping and particle-exclusion process on discrete lattice showed various interesting features, including non-monotonic dependence of velocity on density~\cite{Nishinari2003,D.Chowdhury2010}. 
 However, these models predict jamming at high densities associated with exclusion interaction in a discrete lattice and  random sequential movement of entities governing the dynamics. 
 They fail to capture the absence of jamming in ant traffic as observed in Ref.~\cite{John2009}. Our model takes a different approach. The ants are viewed as particles interacting via nearest-neighbor repulsion. The particles perform continuum dynamics and move together, as opposed to random sequential hopping on discrete lattice considered in Ref.~\cite{Nishinari2003,D.Chowdhury2010}. The biological inputs in the model come through generation of active self-propulsion force in the particles, that has a deterministic part and a stochastic noise.

Most ants have extremely limited eyesight, but are still able to efficiently manage traffic by co-operative trail formation. These trails are made by ants depositing pheromones on the ground which act as signals for the trailing ants to follow the same path. {Pheromones evaporate with a rate dependent on environmental factors. Ants use differential sensing of pheromones to guide their motion. The sensitivity to concentration gradient decreases at high concentrations.  An earlier model of diffusing agents interacting with external field of pheromone, which itself undergoes addition, evaporation and diffusion dynamics led to emergence of trails at reasonable parameter regimes~\cite{Rauch1995}. With time, continuous deposition of pheromones make the signal from a trail strong enough such that all successive ants follow the same path without straying.}  
Our model considers ant motion on preformed trails, thus considering ants as particles moving in one dimension (1D). The motion within this trail could be guided by local sensing -- limited vision or antennal touch. We incorporate a feedback mechanism based on inputs from these local interactions into the active force generation. Our model captures all the main features of experimental results, showing how this feedback can crucially control ant motion. 
We present further predictions that are amenable to experimental verification.

\section{Model and Simulation}
We model the motion of ants in a preformed trail, as one dimensional (1D) system of self propelled  particles (SPP).
The dynamics of $i$-th SPP can be described in terms of the Langevin equations of motion 
\bea
\dot x_i &=& v_i \nn\\
\dot v_i &=&  -\g v_i + \eta_i(t) + F_i -\p_i \sum_{j=i\pm 1} U(x_{ij}),
\label{lange}
\eea 
where $F_i(x_i,x_{i+1})$ is a self-propulsion force, $U(x_{ij})$ denotes a repulsive nearest neighbor interaction ensuring that particles can not cross each other in 1D.  
The viscous dissipation term $-\g v_i$, models dissipation in energy, whose origin may lie within the ant's body - in the movement of muscles that it utilizes to walk, or in friction with local environment, like the walking surface. The  noise $\eta_i(t)$ is interpreted as a stochastic part of self-propulsion, and thus it has a non-equilibrium origin. We assume that the time-scales associated with generation of self-propelled force that comes from an internal energy depot is much faster with respect to the mechanical motion of the ants~\cite{Schweitzer1998}. Thus the stochastic force is assumed to be Gaussian white noise with  
$\la \eta_i(t) \ra =0$, $\la \eta_i(t) \eta_j(t')\ra = 2  D_0 \d(t-t') \d_{ij}$ where $D_0(x_i,x_{i+1})$ denotes non-equilibrium fluctuations.

The interaction potential  between nearest neighbors $U(x_{ij})$ models impenetrability of the ants, with  $x_{ij}=x_j - x_i$ with $j=i\pm 1$. We choose the repulsive part of Lennard- Jonnes potential $U(x_{ij})=4\e[(\s/x_{ij})^{12}-(\s/x_{ij})^6 +1/4]$, with a cutoff distance set to $r_c=2^{1/6}\s$ such that  $U(x_{ij})=0$ if $|x_{ij}| > r_c$. Here $\s$ sets the unit of length and is of the order of the average length of one ant, and $\e$  sets the unit of energy. 
We perform molecular dynamics (MD) simulations using the velocity- Verlet scheme, with integration time step $\d t=0.01 \t$ where 
$\t=\s\sqrt{m/\e}$  is the unit of time and correspond to $1$s.  
We fix the local temperatures at $D_0(x_i,t)/\g \kb$ using Langevin thermostat characterized by an isotropic friction $\g = 1/\t$. 

If one uses a constant self-propulsion force $F_i$, the Langevin model would generate average particle velocity completely independent of density. However, experiments~\cite{John2009} showed a weak but steady decline in velocity with increasing density. This means that the ants sense the local crowding and use a feedback mechanism to control the amount of self-propulsion force generated. Thus we model the self-propulsion force generated by $i$-th ant as $F_i = \g f_0 (1-1/\r_f\d x_i)$
where $\d x_i = x_{i+1}-x_i$, the separation between $i$-th ant and the nearest neighbor in front, and $\r_f$ is a constant. 
In using the distance headway $\d x_i$ to model self-propulsion feedback, we have assumed that the ant senses the position of its leading neighbor using its limited vision or antennal touch, but remains indifferent to the trailing neighbor with regard to self-propulsion force generation.
On an average, $\la \d x_i \ra$ is a measure of inverse local density $1/\r$. 
Using a fit to the experiments on ant-trails~\cite{John2009} we choose $f_0=6.66$ and $\r_f =1.73$ to characterize the force $F_i$ (Fig.\ref{mean_v}).

\begin{figure*}[t]
\begin{center}
\includegraphics[width=8cm]{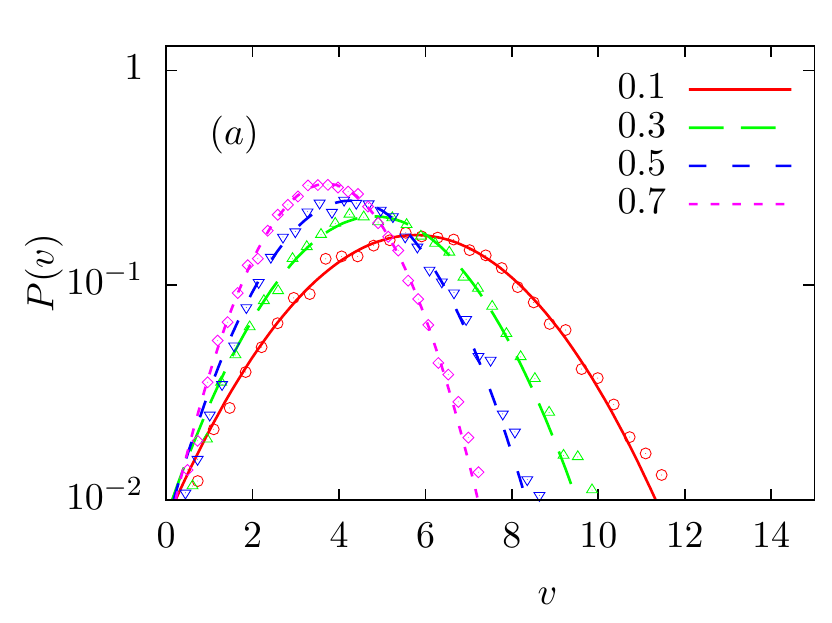}
\includegraphics[width=8cm]{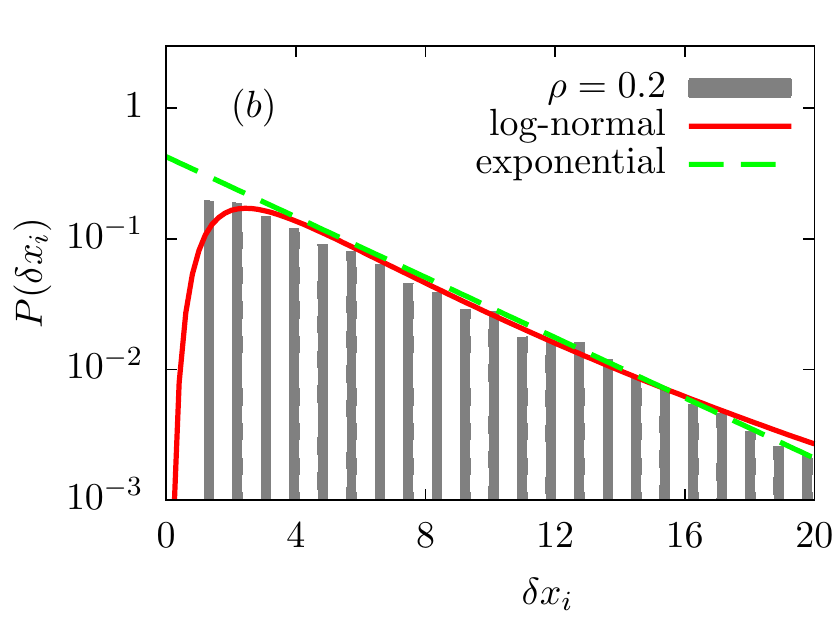}
\includegraphics[width=8cm]{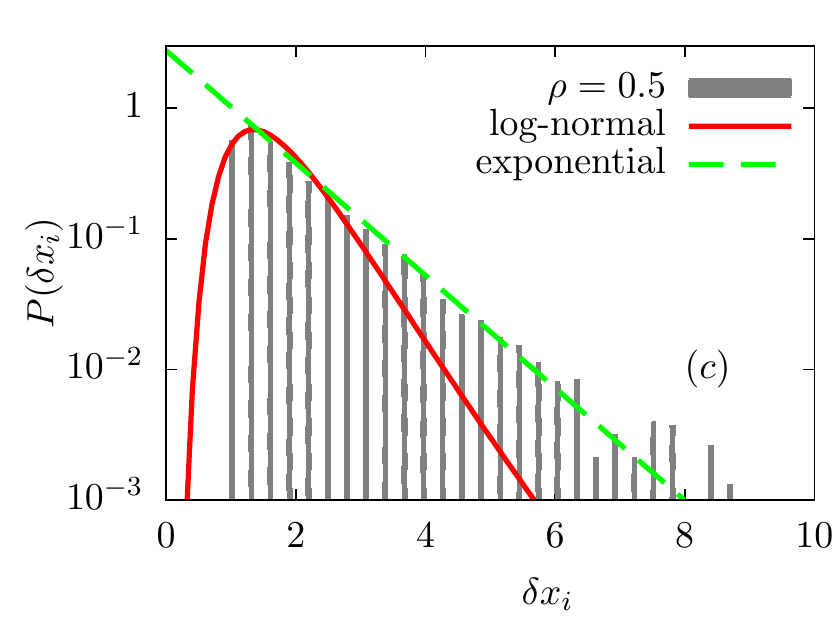}
\includegraphics[width=8cm]{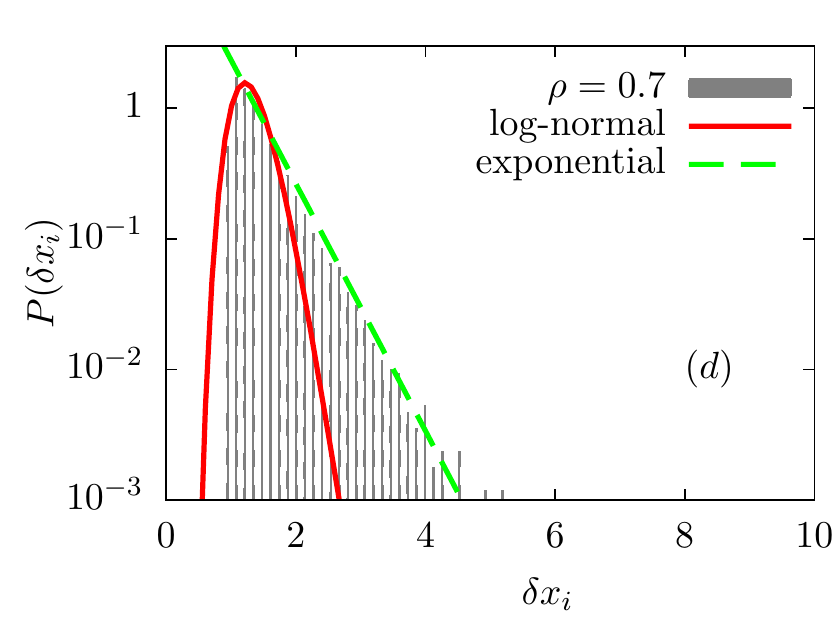}
\caption{(Color online) Distribution functions.
$(a)$~Probability distribution of velocity of each particle at densities $\r=0.1,\,0.3,\,0.5,\,0.7,\,0.9$. The points are obtained from simulation, and the lines show expected Gaussian distributions with varying peaks and widths. 
$(b)$--($d$)~Simulation results for probability distributions of distance headways, at densities $\r=0.2,\,0.5,\,0.7$. }
\label{fig2}
\end{center}
\end{figure*}

The stochastic force $\eta_i(t)$ helps the ants to explore the area around them, e.g., for food, in the absence of external cue like a well formed trail, or odorants from food source. While this noise is a good strategy for exploration, it can be a hindrance in traffic flow once a trail is formed, since it can lead to enhanced collisions. In fact, it is well known from the work by Nagel and Schreckenberg~\cite{Nagel1992} that the introduction of noise in realistic models of traffic leads to traffic jams. We thus model our ants to have a feedback mechanism that reduces noise as the local density increases, leading to reduced collisions and thus reducing the probability of traffic jams.  As for the self-propulsion force above, the simplest such choice would be a linear decrease with local density ($\sim 1/\delta x_i$), i.e. $D_0(x_i,t) = \tilde D_0 (1 - 1/\r_D \d x_i)$ with $\tilde D_0=\g \kb T$ characterizing an equilibrium-like fluctuation strength, and $\r_D$ is a constant. 
However diffusivity $D_0$ has to be positive for all possible $\d x_i$, a condition 
that would be violated  at high densities if the above mentioned linear form were chosen. 
Thus we choose $D_0(x_i,t) = \tilde D_0 \exp(- 1/\r_D \d x_i)$ which captures well the experimentally obtained fluctuations in velocities 
with $\tilde D_0=7.66$ and $\r_D=0.47$, and remains positive at all densities (Fig.\ref{mean_v}).

\begin{figure}[t]
\begin{center}
\includegraphics[width=8cm]{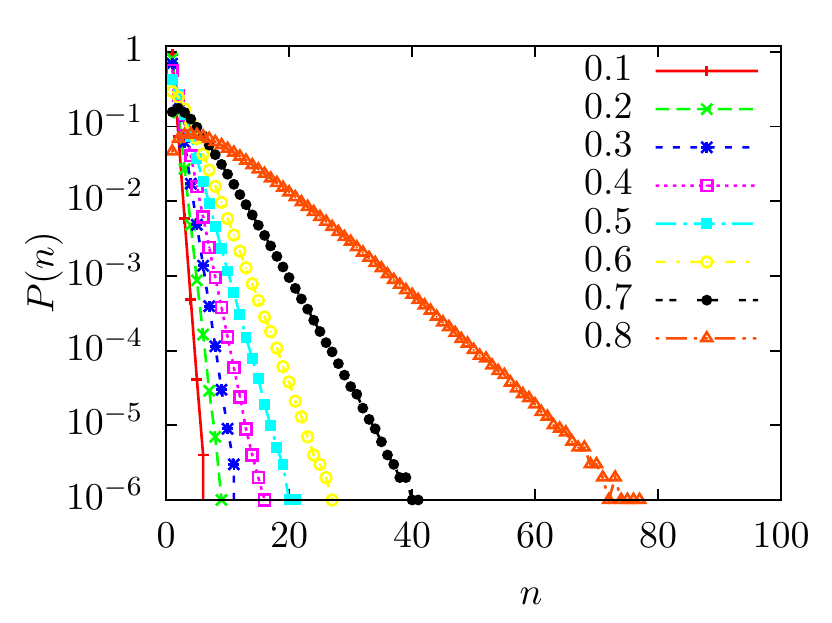}
\caption{(Color online) Cluster- size distributions $P(n)$ at various densities $\r$ denoted in the legend.  The largest possible cluster size is $n=N=4096$. The semi-log plots show exponential tails of the distributions $\exp(-n/n_c)$.
 }
\label{fig3}
\end{center}
\end{figure}

\section{Results and Discussion}

We have chosen our parameters defining the self-propulsion force and fluctuations to fit the data from experiments~\cite{John2009}. 
As can be seen from Fig.~\ref{mean_v}, our results for the mean velocity as well as the variance describe the data well. 
Replacing  $\d x_i$ by the mean-field value $\la \d x_i \ra = 1/\r$, the steady state mean velocity obtained from Eq.~\ref{lange} is $\la v \ra = f_0 (1-\r/\r_f)$, leading to a mean flux $j = \r(1-\r/\r_f)$ which agrees with simulation results. Note that our simple assumption for feedback controlled self-propulsion gives an expression of flux $j$ that has the same behavior as the discrete totally  asymmetric simple exclusion process (TASEP)~\cite{Chou2011}, however with a $\r_f$ that lies at an inaccessibly large value. Thus $\la v \ra$ shows a slight decrease with density in the experimentally accessed regime. 
Unlike the usual traffic model, the current or flow in our system (Fig.~\ref{mean_v}: inset) does not show a congested branch at high densities thus reflecting the absence of jamming.  In Fig.~\ref{fig2}($a$) we show the probability distribution of velocities of individual particles at various values of mean density.  The width of the velocity distribution reduces with increasing density. This happens as the ants reduce the strength of the noise $\eta_i(t)$ in self propulsion using feedback from the local density.  Thus, our model captures the two main features of ant traffic on well-formed trails~\cite{John2009}: absence of jamming even at high densities, and a decrease in velocity fluctuations with increase in density. 

Comparison of our model for ants with Langevin models for traffic~\cite{Mahnke2005a}, shows that the central difference between cars and ants is in the choice of self propulsion force. In traffic models, self propulsion is reduced to zero as the distance between two cars vanishes,  to avoid collision between cars. Whereas in natural ant traffic, ants may come into touching distances. In our model for ants, the active forces decrease but by a small amount as ants approach each other within touching distances ($F_i>0$ at $\delta x_i=1$).  The fact that ants collide with each other is not surprising since they are practically blind and navigate essentially through pheromone sensing.

The fluctuations reveal another important aspect of ant traffic. The reduction of velocity fluctuation with density led to our choice for the diffusion constant getting exponentially suppressed with increase in local density. This ensures that the {\em ant fluid} reduces the local {\em effective temperature} when density increases, to keep a control over the local pressure. This means that while ants do not completely avoid collisions among themselves, they do make sure that the number of collisions per unit time are kept largely unchanged. The reduction of noise strength $D_0$, ensures that at high densities, all ants will generate {\em almost} exactly the same self-propulsion force, thus everybody may move together although being in touching distances. Note that if the noise were independent of local density, a faster ant would stop because of collision with a slower ant -- but if everyone moves with exactly the same velocity, jamming is avoided.

The other quantity that we compare with experimental data is the distribution of headway distances $\d x_i$. Similar to the experiments in Ref.~\cite{John2009}, we find log-normal behavior at short distances $P(\d x_i)=(1/\sqrt{2\pi \s_{\log}^2 \d x_i^2}) \exp[-\{\ln(\d x_i)-\mu\}^2/2 \s_{\log}^2]$ and  exponential behavior at long distances $P(\d x_i)=A \exp (-\d x_i/\l)$ as seen in Figs.~\ref{fig2}(b-d).  To understand the origin of exponential tail in $P(\d x_i)$ we consider the single-file motion.  In a system of 1D hard rods of length $a$, the nearest neighbor distribution at equilibrium is given by  $g_{nn} (x,x') = [\r^2/(1-\r a)] \exp[-(|x-x'|-a)/\l ]$ with $\l=(1-\r a)/\r$~\cite{Corti1998}. Although in our case  the particles are self-propelled, we obtain the same predominantly exponential decay in the distribution of separation between consecutive particles $P(\d x_i)=A \exp (-\d x_i/\l)$ with decay length $\l=(1-\r a)/\r$ that fits well to all the simulation data with $a=0.96$ [Figs.~\ref{fig2}(b-d)]. The origin of the log-normal behavior at small $\d x_i$ is in the non-equilibrium self-propulsion.
We find that the peak in the headway distribution at $\d x_i=1.4$ is independent of the mean system density. This suggests formation of clusters with this typical inter-particle separation within a cluster, irrespective of overall density. Similar behavior was observed in the experiment of Ref.~\cite{John2009}.

In order to probe this point further, we perform a clustering analysis. A randomly chosen particle is assumed to be part of the first cluster. If its nearest neighbors are separated from this particle by a distance less than $1.4$, which is the average headway separation within a cluster, they are also assigned to the same cluster. This procedure is continued until no more particles can be assigned to the first cluster. Then a new random particle which remained unclustered so far is assigned to the next cluster, and the clustering procedure continued in the same way as before until all particles are assigned to a cluster~\cite{Allen1987}. 
The size of clusters $n$ may vary from $1$ to $N$, the total number of particles in the system. The resulting cluster size distributions $P(n)$ calculated from the steady state dynamics of our MD simulation are shown in Fig.~\ref{fig3}. 
In order to obtain better statistics for larger clusters, we performed these simulations using $N=4096$ particles.
The distribution of clusters of ants $P(n) \sim \exp(-n/n_c)$ at all densities $\r<1$, with the tail going up to higher fractions $n/N$ signifying increase in typical cluster size.  
At further higher densities $\r \geq 0.95$ clusters containing all the ants in the system starts to emerge. In the limit of $\r=1$, all the ants belong to the same cluster, as fluctuations of headway distances get completely suppressed.

\begin{figure}[t]
\begin{center}
\includegraphics[width=8cm]{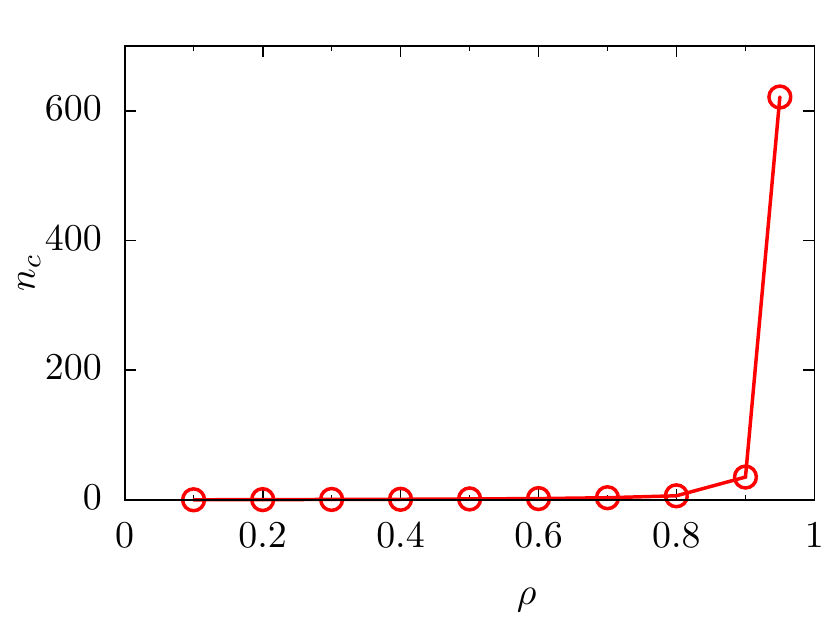}
\caption{(Color online) Typical cluster size $n_c$ as a function of overall density $\r$, shows sharp increase in $n_c$ at $\r > 0.9$, indicating a non-equilibrium first order transition.
 }
\label{fig4}
\end{center}
\end{figure}

We obtain the typical cluster sizes $n_c$ at various densities by fitting $P(n)$ to the exponential form $\exp(-n/n_c)$.  $n_c$ shows a sharp increase for densities $\r > \r_c$ where $\r_c=0.9$ (see Fig.\ref{fig4}). This shows a non-equilibrium first order phase transition towards formation of an {\em infinite cluster}, containing all the ants available in the system. Note that this infinite cluster formation is unlike the aggregation models of Ref.~\cite{Majumdar1998} where the transition was associated with a change in the cluster-size distribution from exponential to power law.

\section{summary \& outlook}
{ We have presented a model for repulsively interacting self-propelled particles undergoing single file motion that shows properties in good agreement with the experimental observations on ants presented in Ref.~\cite{John2009}.} To describe ant-traffic on pre-formed trails, we assumed a generic local crowding dependent feedback control for the deterministic and stochastic parts of self-propulsion force. In agreement with experiments, we find an absence of jamming at all densities. 
Our model captures the decrease in velocity fluctuations observed in real ants, and shows a peak in the headway distribution which is approximately independent of ant density. We performed a clustering analysis to find an exponential cluster size distribution, {  independent of mean density}. The typical cluster size shows a discontinuous increase at a threshold density indicating a first order transition.  These predictions may be verified from further experiments. 

Our model provides a detailed understanding of the dynamics of ants { in preformed trails} and has implications for { technology, e.g., in} 
 mechanisms for self driving cars whose traffic would not jam and robotic swarms that would carry out tasks efficiently and safely like ants. It remains to be seen what patterns emerge from an active walker model with a feedback controlled self-propulsion mechanism proposed in this paper, and whether and to what extent they describe formation of ant trails -- in particular, how far they can describe milling or lane-formation in ants~\cite{Couzin2003}.

\acknowledgements
We thank Debashish Chowdhury of IIT-Kanpur for valuable comments, { and a critical reading of the manuscript}. DC thanks MPI-PKS Dresden for hospitality, where a substantial part of this work was done.

\bibliographystyle{prsty}


\begin{thebibliography}{10}

\bibitem{Vicsek2012}
T. Vicsek and A. Zafeiris, Physics Reports {\bf 517},  71  (2012).

\bibitem{Chou2011}
T. Chou, K. Mallick, and R.~K.~P. Zia, Reports on Progress in Physics {\bf 74},
   116601  (2011).

\bibitem{romanczuk2012active}
P. Romanczuk, M. B{\"a}r, W. Ebeling, B. Lindner, and L. Schimansky-Geier, The
  European Physical Journal Special Topics {\bf 202},  1  (2012).

\bibitem{Hoelldobler1990}
B. Hoelldobler and E.~O. Wilson, {\em The Ants} (Cambridge, Cambridge, 1990).

\bibitem{Dussutour2005}
A. Dussutour, J.-L. Deneubourg, and V. Fourcassi\'{e}, The Journal of
  experimental biology {\bf 208},  2903  (2005).

\bibitem{Dussutour2004}
A. Dussutour, V. Fourcassi\'{e}, D. Helbing, and J. Deneubourg, Nature {\bf
  428},  70  (2004).

\bibitem{Fourcassie2010}
V. Fourcassi\'{e}, A. Dussutour, and J.-L. Deneubourg, The Journal of
  experimental biology {\bf 213},  2357  (2010).

\bibitem{Couzin2003}
I.~D. Couzin and N.~R. Franks, Proceedings of the Royal Society B: Biological
  Sciences {\bf 270},  139  (2003).

\bibitem{Blum2005}
C. Blum, Physics of Life Reviews {\bf 2},  353  (2005).

\bibitem{Dorigo2005}
M. Dorigo and C. Blum, Theoretical Computer Science {\bf 344},  243  (2005).

\bibitem{Ganeshaiah1991}
K.~N. Ganeshaiah and T. Veena, Behav. Ecol. Sociobiol. {\bf 29},  263  (1991).

\bibitem{Millonas1992}
M.~M. Millonas, J. Theor. Biol. {\bf 159},  529  (1992).

\bibitem{Watmough1995}
J. Watmough and L. Edelstein-Keshet, J. Theor. Biol. {\bf 176},  357  (1995).

\bibitem{S.Camazine2001}
S. Camazine, J.~L. Deneubourg, N.~R. Franks, J. Sneyd, G. Theraulaz, and E.
  Bonabeau, {\em Self-Organization in Biological Systems} (Princeton University
  Press, Princeton, NJ, 2001).

\bibitem{Chialvo1995}
D.~R. Chialvo and M.~M. Millonas, The Biology and Technol- ogy of Intelligent
  Autonomous Agents NATO ASI Series, vol. 144  (1995).

\bibitem{Rauch1995}
E. Rauch, M. Millonas, and D. Chialvo, Physics Letters A {\bf 207},  185
  (1995).

\bibitem{Scheidegger1967}
A.~E. Scheidegger, Int. Assoc. Sci. Hydrol. Bull. {\bf 12},  15  (1967).

\bibitem{Helbing1997}
D. Helbing, J. Keltsch, and P. Moln{\'a}r, Nature {\bf 388},  47  (1997).

\bibitem{Helbing1997a}
D. Helbing, F. Schweitzer, and P. Moln{\'a}r, Phys. Rev. E {\bf 56},  2527
  (1997).

\bibitem{Chaudhuri2009c}
D. Chaudhuri, P. Borowski, P.~K. Mohanty, and M. Zapotocky, EPL (Europhysics
  Letters) {\bf 87},  20003  (2009).

\bibitem{Chaudhuri2011b}
D. Chaudhuri, P. Borowski, and M. Zapotocky, Physical Review E {\bf 84},
  021908  (2011).

\bibitem{John2009}
A. John, A. Schadschneider, D. Chowdhury, and K. Nishinari, Physical Review
  Letters {\bf 102},  108001  (2009).

\bibitem{Kerner2009}
B. Kerner, {\em {Introduction to Modern traffic flow: theory and control}}
  (Springer, New York, 2009).

\bibitem{Hodgkin1955}
A.~L. Hodgkin and R. Keynes, The Journal of Physiology {\bf 128},  61  (1955).

\bibitem{Lizana2008}
L. Lizana and T. Ambj{\"o}rnsson, Physical review letters {\bf 100},  200601
  (2008).

\bibitem{Barkai2009}
E. Barkai and R. Silbey, Physical review letters {\bf 102},  050602  (2009).

\bibitem{Hahn1996}
K. Hahn, J. K\"arger, and V. Kukla, Phys. Rev. Lett. {\bf 76},  2762  (1996).

\bibitem{Wei2000}
Q. Wei, C. Bechinger, and P. Leiderer, Science {\bf 287},  625  (2000).

\bibitem{Nishinari2003}
K. Nishinari, D. Chowdhury, and A. Schadschneider, Physical Review E {\bf 67},
  036120  (2003).

\bibitem{D.Chowdhury2010}
D. Chowdhury, K. Nishinari, and A. Schadschneider, {\em Simulating Complex
  Systems by Cellular Automata} (Springer-Verlag, Berlin, 2010).

\bibitem{Schweitzer1998}
F. Schweitzer, W. Ebeling, and B. Tilch, Physical Review Letters {\bf 80},
  5044  (1998).

\bibitem{Nagel1992}
K. Nagel and M. Schreckenberg, J. Phys. I France {\bf 2},  2221  (1992).

\bibitem{Mahnke2005a}
R. Mahnke, J. Kaupu\v{z}s, and J. Tolmacheva,  in {\em Traffic and Granular
  Flow'03}, edited by S. Hoogendoorn, S. Luding, P. Bovy, Schreckenberg, and D.
  M., Wolf (Springer, Berlin, 2005), pp.\ 205--210.

\bibitem{Corti1998}
D. Corti and P. Debenedetti, Physical Review E {\bf 57},  4211  (1998).

\bibitem{Allen1987}
M.~P. Allen and D.~J. Tildesley, {\em Computer simulation of liquids} (Oxford
  University Press, New York, 1987).

\bibitem{Majumdar1998}
S. Majumdar, S. Krishnamurthy, and M. Barma, Physical Review Letters {\bf 81},
  3691  (1998).

\end{thebibliography}

\end{document}